\documentstyle[epsfig, graphics]{article}


\newcommand{\be}{\begin{equation}}
\newcommand{\ee}{\end{equation}}
\def\mynote#1{}

\def\qcdm{$Q$CDM}
\newcommand{\eff}{{\sf F}}

\def\qfield{$Q$-field }
\def\pri#1{#1^{\prime}}
\def\qfrac{{\delta\rho_Q \over \rho_Q}}
\def\mfrac{{\delta\rho_m \over \rho_m}}

\newcommand{\unitm}{{\sl m}}           

\def\mxth{\mathsurround=0pt }
\def\xversim#1#2{\lower2.pt\vbox{\baselineskip0pt \lineskip-.2pt
    \ialign{$\mxth#1\hfil##\hfil$\crcr#2\crcr\sim\crcr}}}
\def\ltsim{\mathrel{\mathpalette\xversim <}}

\newcommand{\deebee}{{{\delta \rho_B} \over \rho_B}}
\newcommand{\bee}{(\log{a})} 
\newcommand{\muu}{\mu}
\begin{document}

\title{Sensitivity of the Cosmic 
Microwave Background Anisotropy to Initial Conditions in
Quintessence Cosmology}

\author{Rahul Dave\footnote{Department of Physics \& Astronomy, Univ. of
Pennsylvania, Philadelphia, PA 19104}, R. R. Caldwell\footnote{Department of
Physics \& Astronomy, Dartmouth College, Hanover, NH 03755}, and Paul J.
Steinhardt\footnote{Department of Physics, Princeton University, Princeton, NJ
08544}}
\maketitle
\begin{abstract}
We analyze the evolution of energy density fluctuations in
cosmological scenarios with a mixture of cold dark matter and
quintessence, in which the quintessence field is modeled by a constant
equation of state. We obtain analytic expressions for the time
evolution of the quintessence perturbations in models with light
fields.  The fluctuations behave analogously to a driven harmonic
oscillator, where the driving term arises from the inhomogeneities in
the surrounding cosmological fluid.  We demonstrate that the
homogeneous solution, determined by the initial conditions, is
completely sub-dominant to the inhomogeneous solution for physically
realistic scenarios. Thus we show that the cosmic microwave background
anisotropy predicted for such models is highly insensitive to the
initial conditions in the quintessence field.
\end{abstract}
\section{Introduction}
\label{secintro}
\mynote{WHY Q,WHATS Q}
Growing
observational evidence suggests that the total matter density of
the universe is significantly less than the critical density.
Yet, measurements of the cosmic microwave background (CMB) indicate that the
universe is flat and the total energy density is precisely equal to the critical
density \cite{Ostriker}. Candidates for the missing energy are the 
cosmological constant ($\Lambda$) \cite{LambdaRef}, 
and quintessence ($Q$), a time-varying, 
spatially inhomogeneous component with negative pressure
\cite{PJS,Cald98}. Examples of quintessence include slowly evolving fields 
or topological defects such as a network of light and tangled cosmic
strings \cite{Spergel}-\cite{kess}. Cold Dark Matter (CDM) models with 
quintessence (\qcdm) which fit the data from
observations of high red shift supernovas, gravitational lensing, 
CMB anisotropy,  and structure  formation \cite{Ostriker}-\cite{PJS}
have been found.

\mynote{WHICH MODELS,WHAT'S INSENSITIVITY?}
For the purposes of this paper, we model quintessence
as a scalar field quintessence rolling down a
potential $V(Q)$ with an equation of state $w \equiv p_Q / \rho_Q$, where $p_Q$ is the
pressure and $\rho_Q$ is the energy density of the field.
We consider here a large class of models in which the field and
fluctuation evolution may be represented by a constant equation of state 
with value between 0 and -1, and in which the sound speed in quintessence fluctuations, 
$c_{sQ}^2$, approaches unity at scales much smaller than the horizon. 
Most models of quintessence 
that have appeared in the literature satisfy these 
conditions. This class includes not only ``tracker models'', which 
have dynamical attractor behavior, but also more 
general potentials \cite{Weiss}-\cite{zlatev}.
We show that, in such models, 
the CMB anisotropy is insensitive to initial conditions of the
quintessence field.
By insensitive we mean that the fractional change in the CMB anisotropy 
power spectrum due to a change in initial conditions cannot be 
observationally resolved. 

These conclusions were originally mentioned in our first paper on quintessence
models \cite{Cald98}, and also by others 
\cite{Viana:1997mt, Perrottaetal, Braxetal}, 
who
noted that the amplitude of the energy density perturbations in the
background fluid are largely independent of the initial conditions in
the scalar field, provided the initial energy contrast is less than
unity. Our present work is consistent with these earlier
conclusions. However, we go further in this paper by examining
in detail the behavior of the scalar field
perturbations, and the causes for the insensitivity, as described next.

\mynote{ROADMAP}In the next section we derive the equations of motion for
the evolution of the \qfield and its fluctuations. We introduce a
formalism to numerically study the 
quintessence fluctuations in terms of the evolution of the 
equation of state as a function of cosmological scale-factor. 
In section \ref{section.approx} we show that models with constant equation
of state represent the behavior of a large
class of models with light fields and monotonically evolving equations of state. 
In section \ref{section.evol} we analytically solve the fluctuation equation 
for constant equation of state at large wavelengths. We use this solution to 
describe key features of the fluctuation evolution obtained from the numerical integration.
Finally, in section \ref{secinits} we report the effect
of changing initial conditions on quintessence fluctuation evolution and
CMB anisotropy. We show that the anisotropy is highly insensitive to changes 
in the initial conditions. 
\mynote{For example, consider models with $w > -0.9$ and long wavelength matter
fluctuations with amplitude $\delta \rho_m / \rho_m \sim 10^{-5}$ at 
horizon re-entry (which corresponds to an an initial value of
$\delta \rho_m / \rho_m \sim 10^{-16}$ at $a=10^{-8}$ in the synchronous gauge).
Then, for any initial value of $\delta \rho_Q/\rho_Q$ less than $10^{11}$ in
the synchronous gauge, there
is no distinguishable difference in the CMB anisotropy.}

\section{Field and fluctuation equations}
\label{secqeqn}

\mynote{SETUP, FIELD EQUATIONS}
We consider a matter-quintessence Lagrangian of the form  
\begin{equation}
\label{eqn.lag}
L = L_{B} + L_{Q},  
\end{equation}
where the {\em B} refers to all the background species of particles 
and fields, including baryons, photons, cold dark matter, and neutrinos.
The background cosmology is described by the
Friedman-Robertson-Walker metric with positive signature.
We model the $Q$-field as a classical,
self-interacting, scalar field, minimally coupled to other constituents
of the universe through gravity,
\begin{equation}
L_{Q} = {1 \over 2} Q_{,\mu} Q^{,\mu} - V(Q).
\end{equation}
We only consider models with canonical kinetic energy terms, since
$c_{sQ}^2$ approaches unity at scales much smaller than the horizon \cite{Cald98} in these
models.

The time evolution of the quintessence field is determined by the equation of motion
\begin{equation}
\label{eqn.bg}
Q'' + 2 {a' \over a} Q' + a^2 {\partial V \over \partial Q} = 0,
\end{equation}
where $a$ is the scale-factor, normalized to unity today, 
and where the prime ($'$) represents $\partial/\partial\tau$, the
derivative with respect to conformal time. 
By specifying the functional form of the potential, $V(Q)$, along with initial
conditions $Q,\, Q'$ at a time $\tau_{init}$, the subsequent evolution is
determined for all times $\tau > \tau_{init}$.

At the scales of interest 
for the cosmic microwave background ($ > 1 Mpc$),
fluctuation amplitudes of the $Q$-field and the metric are 
usually small
compared to the fields themselves, and thus a linearization of the
$Q$-field and Einstein  equations in the perturbation suffices to
describe the fluctuation dynamics.
Our analysis, carried out in the synchronous gauge, 
uses the conventions and equations 
from Ma and Bertschinger \cite{MaandB}.

\mynote{FLUCT EQNS}
The synchronous gauge is defined by the condition that the time-time
and time-space
part of the metric are not perturbed.
The perturbed metric is given as:
\begin{equation}
  ds^2 = a^2(\tau)\{-d\tau^2 + (\delta_{ij} + h_{ij})dx^i dx^j\}\,.
\end{equation}
The metric perturbation $h_{ij}$ can be decomposed into a trace part
$h \equiv h_{ii}$ and a traceless part $\eta$. 
To enable us to
study fluctuations as a function of wavelength, we will work in
Fourier space ($\vec{k}$) with perturbations 
$h(\vec{k},\tau)$ and $\eta(\vec{k},\tau)$.
         
The time evolution of the perturbed metric is obtained by linearizing
the Einstein Equations in Fourier space (see \cite{MaandB}). 
The cosmological perturbation equations for the dynamical $Q$-component are
obtained by expanding the scalar field equations about the homogeneous
background. In the synchronous gauge, we can write down the equation for 
small fluctuations $\delta Q$ in Fourier space:
\be
\delta Q'' + 2 {a' \over a} \delta Q' + (a^2 {\partial^2 V \over \partial Q^2}
+ k^2) \delta Q = -{1 \over 2} h(\vec{k},\tau)' Q'.
\label{deltaQeqn-ink}
\ee
The fluctuation amplitude evolves in time like a
scalar field.

To compute the quintessence fluctuation evolution and CMB anisotropy power spectrum,
we use the fluctuation and Einstein equations to evolve the
moments of the photon distribution in a Boltzmann code.
We modified two separate Boltzmann codes,
 CMBFAST (\cite{Sel})  and  LINGER (\cite{MaandB}), by 
adding in scalar field evolution, and repeated 
all computations in the conformal gauge. The results from these multiple approaches
were all identical to better than 1 part in $10^8$ \cite{THESIS}.

\mynote{EQN OF STATE PARAMETERIZATION}
We can parameterize quintessence models in terms of the evolution of the equation of 
state as a function of scale-factor $w(a)$. The 
formulation is useful in studying the time dependence 
of quintessence fluctuations and in unearthing the similarities in time evolution 
of models with different potentials. The energy density and pressure of the
$Q$-component in this parameterization are given by
\begin{equation}
\rho_Q(a) = {3 H^2_0 \Omega_Q \over 8 \pi G}  
\exp{\Big( 3 \big[-\log a 
+ \int^{1}_a {d \tilde a \over \tilde a} \, w(\tilde a) \big] \Big)},
\label{rhoQeqn}
\end{equation}
and 
\begin{equation}
\qquad p_Q(a) = w(a) \rho_Q(a),
\label{pQeqn}
\end{equation}
where $H_0$ is the present-day value of 
the Hubble constant, and $\Omega_Q$ 
is the present-day $Q$-field energy
density as a fraction of the critical energy density.

To obtain the equations of motion in terms of the equation of state, we express the first 
and second derivative of the potential in terms of $w$:
\begin{eqnarray}
\label{eqn.explicit}
a^2 V_{,Q} &=& -{1 \over 2}  
\Big( 3 {a' \over a} (1 - w )  + {w' \over 1 + w } \Big) Q',  \\
a^2 V_{,QQ} &=& 
-{3 \over 2}(1-w )
\Big[ {a'' \over a} - \Big({a' \over a}\Big)^2\Big({7 \over 2}
+ {3 \over 2} w \Big) \Big] \cr\cr
&& + {1 \over 1 + w }\Big[ {w'^2 \over 4(1 + w)} - {w'' \over 2}  
+ w' {a' \over a}(3 w + 2)\Big] .
\label{eqn.explicit2}
\end{eqnarray}

If we define $\delta\psi = \delta Q/\sqrt{1 + w(\tau)}$ and 
$\psi'=Q'/\sqrt{1 + w(\tau)}$, we can use Eqs. \ref{eqn.explicit}
and \ref{eqn.explicit2} 
 to convert the field (Eq. \ref{eqn.bg}) and fluctuation 
(Eq. \ref{deltaQeqn-ink}) equations for the $Q$-field 
 into a form in which the potential is implicit:
\begin{equation}
\label{invertedfieldeqn}
\psi'' + \Big((1 + 3w) {a' \over a} \Big)\frac{\psi'}{2} = 0,
\end{equation}
and 
\begin{eqnarray}
\label{combinedflucteqn}
{\delta\psi'' + \Big(2 {a' \over a} + {w' \over 1 + w}\Big)\delta\psi'}  \nonumber & \\
+ \Big(k^2
-{3 \over 2}(1-w)\Big[{a'' \over a} - \Big({a' \over a}\Big)^2\Big({7 \over 2}
+ {3 \over 2} w\Big) \Big]
+ 3 w'{a' \over a}\Big)\delta\psi \nonumber & \\ 
=  - {1 \over 2} h' \psi' 
\end{eqnarray}
A practical consequence of this change of variables is that w''
drops out of the evolution equation, and we only need to specify w
and w' as functions of time.
Thus, given a well sampled table of the equation of state history $w(a)$, we can 
numerically integrate this equation in a Boltzmann code to obtain 
the evolution of the quintessence field and its fluctuations.
\section{Why constant $w$ approximates well a large range of potentials}
\label{section.approx}

\mynote{MONOTONICITY}
For quintessence models described by a scalar field with potential $V(Q)$, 
the equation of state varies with the scale-factor, depending on the initial 
conditions on $Q$, and the form of $V(Q)$. 
In certain cases, especially for large mass fields (e.g. harmonic potential with large mass field,
$m >> H_0$), the evolution is oscillatory.
In most cases, however, the mass of the quintessence field 
is smaller than or comparable to the Hubble parameter ($m \ltsim H_0$), and 
the evolution of $w(a)$ is monotonic. Observations
are consistent with models in which the 
equation of state evolution is monotonic and slowly varying \cite{Ferreira,zlatev}.
\mynote{, where $w(a)$ is either decreasing
from 0 to -1, or increasing from -1 to 0 }

\begin{figure}[h]
\centerline{\epsfxsize=5 in \epsfbox{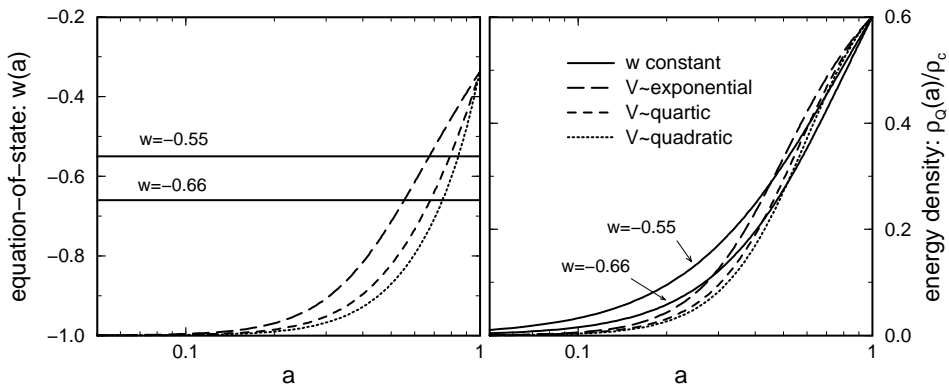}}
\caption{
In the left panel, we see the evolution of the equation of state for a set 
of potentials, all with $\Omega_Q=0.6$ and $w=-1/3$ today. 
The evolution of the ratio of the quintessence 
energy density to the critical energy density is shown in the right panel.
In both panels, the upper, solid 
curve represents a constant $w=-0.55$ model, while the lower solid curve
represents a constant $w=-0.66$ model.
The $w=-0.55$ model has been chosen to best-fit approximate the
exponential potential, while the $w=-0.66$ model has been
chosen to best-fit approximate the quadratic and quartic  potentials. 
Notice the similarity in the energy density evolutions.
}
\label{whyapprox}
\end{figure}

\mynote{APPROXIMATING WITH CONSTANT $w$}\mynote{ENERGY DENSITY HISTORY SIMILARITY}
 In Figure \ref{whyapprox} we consider  
equation of state histories in a group of models with different potentials and
initial conditions such that $w(a)$ increases monotonically.
The examples plotted
are quadratic, quartic, and exponential potentials \cite{ExpoPot}.
We compare the evolution of the cosmological energy density in quintessence, $\rho_{Q}(a)$, 
in these models with that from models with constant equation of state.
The parameters and initial conditions have been chosen so that each case
produces the same present-day values for the 
equation of state and total energy density. The left 
panel shows the evolution of the equation of state, while the right panel shows
the evolution of the energy density. In both panels, the upper, solid 
curve represents a constant $w=-0.55$ model, while the lower solid curve
represents a constant $w=-0.66$ model. The $w=-0.55$ model has been chosen to best-fit approximate
the exponential potential, while the 
$w=-0.66$ model has been chosen to best-fit approximate the quadratic and quartic potentials. 
For each of the potentials, $\rho_{Q}(a)$ is a 
monotonically increasing function of the scale-factor. 
Although the time-history of the equation of state is quite different between 
the constant $w$ and evolving
potential cases, we can see from the figure that 
the evolution of the energy densities is closely comparable.

Quintessence affects the CMB anisotropy chiefly through effects which 
depend on $\rho_Q(a)$, that change the expansion history of the universe \cite{Cald98}.
Thus, these potentials predict nearly identical CMB 
anisotropy to the best-fit constant $w$ models with 
$w = w_{eff}$. The effective equation of state is 
empirically obtained as  the
$\rho_Q$ weighted average value of $w(a)$ \cite{THESIS}:
\begin{equation}
w_{eff} = \frac{\int da \, w(a) \, \rho_Q(a) \,}{\, \int da \, \rho_Q(a)} .
\label{weffective}
\end{equation}

\begin{figure}[h]
\centerline{\epsfxsize=2.8 in \epsfbox{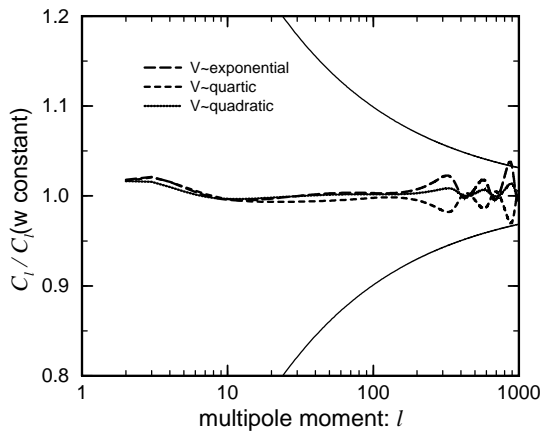}}
\caption{
 Here we plot the ratio of the
power spectrum in the models from Figure \ref{whyapprox} to the power spectrum
in the corresponding best-fit constant $w$ model. The fractional cosmic
variance with respect to the best-fit model is also shown (outer thin lines). 
The ratio for each model falls well 
within this variance envelope at most of the
multipole moments, thus, the predicted anisotropy is
observationally indistinguishable.
}
\label{fig39}
\end{figure}

\mynote{CMB CONSEQUENCE}
In Figure \ref{fig39} we plot the ratio of the
multipole moments of the CMB power spectrum in the
exponential, quartic, and quadratic potential models
to the power spectrum of the corresponding model with constant 
equation of state $w=w_{eff}$. Also plotted is the fractional cosmic variance.
 (Cosmic variance is the intrinsic theoretical uncertainty
for any model prediction based on adiabatic gaussian perturbations.)
The ratio in each of the cases falls within the cosmic variance
uncertainty for almost 
all multipoles. Hence, the power spectrum in each case is 
observationally indistinguishable from the corresponding constant $w$ model.

\mynote{RESTRICTION}
Thus, for a large class of models with monotonically changing $w$, 
the evolution of quintessence and its fluctuations are described, to within cosmic variance, 
by a constant effective equation of state. 
In this paper we restrict ourselves to these models, and additionally require 
the sound speed in the quintessence fluctuations, or the group velocity of the fluctuations, 
$c_s^2$ to be $\sim 1$ at sub-horizon scales. We do not deal with oscillatory
equations of state or models in which the kinetic energy is 
non-canonical and $c_s^2 < 1$ at smaller wavelengths, 
such as k-essence models \cite{Erik}.

\section{Solving fluctuation equations for constant $w$}
\label{section.evol}

\begin{figure}[h]
\centerline{\epsfxsize=2.9 in \epsfbox{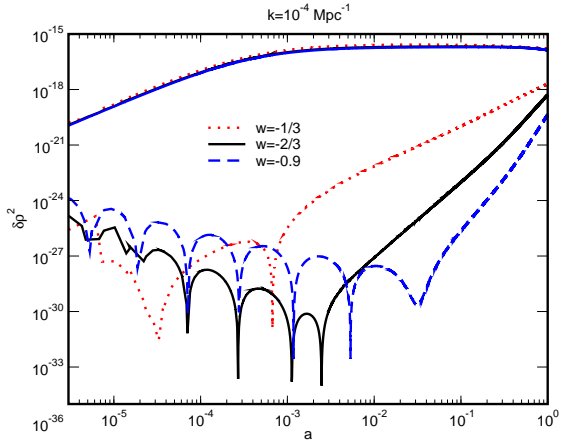}}
\caption{
We compare the evolution of quintessence and matter
fluctuation energy density for a long wavelength $k=10^{-4} Mpc^{-1}$ mode 
in three different models with equations of state $w=-1/3$, $w=-2/3$, 
and $w=-0.9$, and with $\Omega_Q=0.6$, $h=0.65$, and $\Omega_Bh^2 = 0.02$.
The three lower curves are the quintessence fluctuation evolutions at the different 
equations of state, while
the three upper curves, all very close to each other, are the corresponding matter fluctuation
evolutions. Notice that the energy density in quintessence fluctuations changes with equation
of state, but remains much smaller than the energy density in matter fluctuations.
}
\label{k4comp}
\end{figure}

\mynote{FEATURES OF FLUCT EVOLUTION}
To study the evolution of quintessence fluctuations, 
we numerically evolve the fluctuation and Einstein equations 
in a $Q$CDM model with $\Omega_Q = 0.6$, $h=0.65$, and $\Omega_Bh^2 = 0.02$.
In Figure \ref{k4comp} we plot the quintessence and matter fluctuation 
energy density obtained in this model 
for a mode with wavelength larger than the horizon today ($k=10^{-4} Mpc^{-1}$).
The lower three curves are the fluctuation evolutions at three different
equations of state, $w=-1/3$, $w=-2/3$ and $w=-0.9$. 
We see in the figure that for all the equations of state, the fluctuation amplitude first 
oscillates and decreases, reaches a minimum, and then
ultimately starts to increase. The decrease of the amplitude is sustained for
a longer time, and the subsequent increase is sharper, as
$w$ becomes more negative, i.e., closer to -1. The upper three curves, which are all
almost on top of each other, are the corresponding matter fluctuation 
evolutions. The change in the quintessence fluctuation evolution as a function
of $w$ does not impact the matter fluctuation evolution at all.

\mynote{WORK IS NUMERICAL, BUT WE'RE GONNA APPROXIMATELY EXPLAIN IT}
To understand the nature of this evolution, we look for analytical solutions to the 
quintessence fluctuation equation. We collect together the other constituents of the 
universe into an adiabatic background fluid denoted by the label `$B$', 
with equation of state $w_B$ and energy density $\rho_B$. 
We can then solve the fluctuation equation analytically in two limits: (a) the energy density
in the quintessence field is negligible compared to that in the other 
constituents of the universe ($\rho_Q << \rho_B$), and (b) the energy density in quintessence
dominates the energy density in the rest of the constituents ($\rho_Q >> \rho_B$).
In these limits we can solve for $\delta\psi$
deep inside the radiation- and matter- dominated epochs where $w_B$ is a constant,
and deep inside the quintessence-dominated epoch respectively. In both cases,
the fluctuation equation (Eqn. \ref{combinedflucteqn}) then simplifies to the form 
of a forced harmonic oscillator with constant
coefficients and a force term dictated by the coupling of quintessence to 
the perturbed metric:
\begin{equation}
\label{eqn.bfl}
  {d^2 \delta\psi \over d\bee^2} + \nu {d \delta\psi \over d\bee}  + 
  (k^2 + \unitm^2) \delta\psi =
  - {1 \over 2} \ {dh \over d\bee} { d\psi \over d\bee}.
\end{equation}
Here $k$ is the co-moving wave number,
 \begin{equation}
  \nu = {3 \over 2} (1 - w_B)
  \end{equation}
is the damping coefficient of the oscillator equation, and
\be
\unitm = {3 \over 2} \sqrt{(1 - w) (2+w+w_B)}
\ee
is the Compton mass of the field $\delta\psi$, all expressed in units of the 
co-moving Hubble parameter (${a' \over a}$).

\mynote{SOLUTIONS}
There are two qualitatively different types of solutions to this
fluctuation equation.
The homogeneous solutions
belong to a two-parameter family specified by the initial conditions 
$\delta \psi_{init}$ and $\, \delta \psi'_{init}$, and are
 unaffected by the fluctuations in the background cosmological fluid. The inhomogeneous
solutions, on the other hand, arise as the response 
of the quintessence field to the fluctuations of the 
background. The evolution of quintessence 
fluctuations as a function of scale-factor is determined 
by the combination of these solutions. The
solutions are different in the radiation-, matter- and $Q$-dominated epochs, and  
must be matched by continuity at the boundary between 
radiation and matter domination, and between matter and quintessence domination. 


\mynote{SOLVING INHOMOGENEOUS EQN}
Solving the inhomogeneous equation
requires knowledge of the time evolutions of the 
quintessence field and the metric perturbations. The former evolution can be 
obtained from the equation of motion for the field (Eqn. \ref{invertedfieldeqn}):
\begin{equation}
\label{eqn.psider}
{ d\psi \over d\bee} = 
\sqrt{ {3 H^2_0 \Omega_Q\over 8 \pi G}} {a^2 \over a'} \Big({1 \over a}\Big)^{{3 \over 2}(1+w)},
\end{equation}
while the latter dependence can be obtained from the Einstein equations \cite{MaandB}:
\begin{equation}
\label{eqn.metder}
  {d^2h \over d\bee^2} + {(1 - 3w_B) \over 2} {dh \over d\bee} 
  = -3 ({{\delta\rho + \delta p} \over \rho}) 
  \simeq -3 \deebee (1 + w_B),
\end{equation}
where $\deebee$ is the fractional energy density in the background fluctuations.
The behavior of $\deebee$ at scales larger than the Jean's length,  
the largest scale at which the collapse of a fluctuation through
gravitational instability can be counteracted by the propagation of 
mechanical disturbances in the baryon-photon fluid, in the synchronous gauge, 
is well known \cite{Crossroads}:
\begin{equation}
\label{eqn.thedbee}
  \deebee =  \Big(\deebee\Big)_{H_{I}}\,\Big({a \over a_{H_{I}}}\Big)^{{p \over 2}}.
\end{equation}
Here $\Big(\deebee\Big)_{H_I}$ is the fractional
background energy density and $a_{H_{I}}$ is the scale-factor 
when the fluctuation mode under consideration exits the horizon during
inflation ($H_I$). The density power spectrum at horizon crossing is taken to be a 
scale invariant Harrison-Zeldovich spectrum with a COBE normalized amplitude $D_B$,
\be
\Big(\deebee\Big)_{H_{I}} = D_B k^{{n -1} \over 2}, n=1,
\ee
with $p=4$ deep in the radiation dominated
epoch, and $p=2$ deep in the matter dominated one \cite{Ratra}.

\mynote{LONG WAVELENGTH INHOMOGENEOUS SOLN, $w$ AND $a$ DEPENDENCE}
We use Eqn. \ref{eqn.psider} and the solution of Eqn. \ref{eqn.metder} to 
solve the fluctuation equation (Eqn. \ref{eqn.bfl}) for the evolution of metric 
perturbations, and consequently for the
inhomogeneous solution in the radiation and matter dominated epochs. At 
wavelengths much larger than the Jean's length, we find that:
\begin{equation}
\label{eqn.ihom}
  \delta \psi_I  = - c_I \,D_B\,a_{H_{I}}^{-2}\,a^{{\muu \over 2}},
\end{equation}
where $\muu = p + 3(w_B - w)$, and 
\begin{equation}
\label{eqn.ri}
  c_I = 6\,\Big({\Omega_R \over \Omega_M}\Big)^{2 -{p \over 2}}\,\Big(\sqrt{\frac{\Omega_Q}{\Omega_B}}\,\Big)\,\Big(\sqrt{\frac{\rho_c}{H_0^2}}\,\Big)\,\frac {1} {\muu^2 + 2\muu\nu + 4\unitm^2}.
\end{equation}
Notice that the inhomogeneous solution depends on two separate epochs, $a_{H_I}$, and 
$a_0 =1$, the latter entering the equation through the dependence on the evolution of the quintessence
field.

The magnitude of the inhomogeneous solution is proportional to the amplitude $D_B$ of
the Harrison-Zeldovich spectrum and is thus determined by the COBE normalization.
Since $\muu > 0$, the solution increases with 
increasing scale-factor for all equations of state. This behavior corresponds to the
gravitational amplification of large wavelength quintessence fluctuations 
due to CDM potential wells.
Furthermore, the inhomogeneous solution at a given scale-factor 
is smaller for values of $w$ closer to -1, since the $a^{\mu \over 2}$ scaling
and coefficient $c_I$ are both smaller for more negative values of $w$ and
the $a_{H_{I}}^{-2}$  term is almost independent of $w$.
The reduced amplitude reflects the smaller
coupling to the background in the source term of the fluctuation 
equation (Eq. \ref{deltaQeqn-ink}).


\mynote{HOMOGENEOUS SOLN}
The homogeneous equation has the form of a 
damped harmonic oscillator with constant coefficients. Thus the
solution at all wavelengths in each epoch is simply:
\begin{equation}
\label{eqn.hom}
  \delta \psi_H = c_H a^{-{\nu \over 2} } \theta(a,k,\unitm,\nu) ,
\end{equation}
where $c_H$ is the amplitude of the solution
and where $\theta$ is an oscillatory function of order unity. 
 Since $\nu > 0$ in all epochs, the oscillation envelope decreases as
a power law of the scale-factor.
The amplitude $c_H$ must be determined by the initial conditions on the quintessence 
fluctuations at the initial hyper-surface far outside the horizon, 
deep in the radiation dominated epoch (we chose $a_{init} \sim 10^{-8}$ in our simulations 
and analysis). 

\mynote{TO OBTAIN $c_H$, CONSIDER ADIABATIC INITS}
 The inhomogeneous solution 
scales as a positive power of $a$, and is hence negligible
 ($< 10^{-20}$) at the initial hyper-surface.
 The initial fluctuations in quintessence are thus entirely due to the 
homogeneous solution ($\delta\psi_{init} = \delta\psi_{H,init}$). 
Hence, a change in initial conditions affects the homogeneous solution only. 
To determine $c_H$, we consider the initial conditions predicted by inflation.
Inflation creates a nearly scale-invariant primordial spectrum of
adiabatic density perturbations in all light fields.
Since the quintessence fields of interest in this paper are also 
light fields, the entropy perturbation for the entire fluid, just after inflation, vanishes:
\begin{equation}
T \delta S = \delta p - {p' \over \rho'} \delta \rho = 0.
\label{ic1}
\end{equation}
This condition gives one equation between the initial fluctuations $\delta\psi_{init}$
 and $\delta\psi'_{init}$. A second constraint
is obtained from the observation that long wavelength fluctuation modes are frozen 
outside the horizon, and thus we set:
\be
\label{ic2}
\delta\psi'_{init} = 0.
\ee

\mynote{EXPRESSION FOR $c_H$, $w$ AND $a$ DEPENDENCE OF HOM. SOLN}
We solve the constraint equations for the amplitude of the homogeneous solution:
\be
\label{eqn.chw}
c_H = -{1 \over 6}\,\Big(\sqrt{\frac{\rho_c}{H_0^2}}\,\Big){\Omega_M \over \sqrt{\Omega_Q \Omega_R}}\,{{{a_{init}^{{6+3w} \over 2}}a_{H_{I}}^{-2}} \over {1 - w^2}}\,D_B.
\ee
The declining power law scaling ($a^{-{\nu \over 2}}$) of 
the homogeneous solution is independent of $w$. 
Thus, the equation of state dependence of the 
homogeneous solution comes only from its amplitude $c_H$.
Consequently, the value of the homogeneous solution at a given scale-factor is 
larger for $w$ closer to -1.

\mynote{EVOLUTION OF LONG WAVELENGTH MODES}
Having obtained the approximate solutions of the fluctuation equation
as a function of scale-factor and equation of state (Eqns. \ref{eqn.ihom} and \ref{eqn.hom}), it 
is now possible to understand the numerically obtained long wavelength evolution
shown in Figure \ref{k4comp}. Firstly, note that the scale of both the solutions
is determined by the amplitude of the matter fluctuation at horizon re-entry, and consequently
the COBE normalization. Secondly, we can see from the equations that that the homogeneous
solution decreases as $a^{-{\nu \over 2}}$, while the inhomogeneous solution 
increases as $a^{{\mu \over 2}}$. 
Thus the amplitude of the fluctuations decreases until the
inhomogeneous solution becomes comparable to the homogeneous solution, and 
then it starts to increase. 
The scale factor at which the solutions become comparable, 
\be
\label{aturn}
a_T = \Big({-c_H {a_{H_{I}}}^{2} \over {c_I\,D_B}}\Big)^{4 \over {7 - 6w}},
\ee
increases from $a_T \sim 10^{-5}$ at $w=-1/3$ to $a_T \sim 1.7 \times 10^{-3}$ at $w=-2/3$ to
$a_T \sim 4 \times 10^{-2}$ at $w=-0.9$. 
Thus, as can be seen in the figure, 
the homogeneous solution is comparable to the inhomogeneous one 
for the $w=-1/3$ model at last scattering ($a \sim 7 \times 10^{-4}$), while 
it dominates the inhomogeneous solution
at both the more negative equations of state, $w=-2/3$ and $w=-0.9$.

\mynote{$w$ DEPENDENCE OF EVOLUTION}
The magnitude of the  homogeneous solution at a given value
of the scale factor increases as $w$ approaches -1.
By contrast, the magnitude of the inhomogeneous solution decreases.
Additionally, since 
\be
\delta\rho_Q \propto \psi' \propto {1 \over a^{{1+3w \over 2}}},
\ee
 the energy 
density in the homogeneous solution at a given scale-factor 
further increases with decreasing $w$ for all $w < -1/3$ \cite{THESIS}.
Consequently, in models with $w$ closer to -1, such as $w=-0.9$, 
the energy density of the quintessence fluctuations 
is initially larger and decreases more gradually. 
Thus, the evolution
remains dominated by the homogeneous solution until a later time.

\mynote{SMALLER WAVELENGTHS}
The power law decline of the
homogeneous solution is independent of wavelength.
On the other hand, the amplitude of the inhomogeneous solution for
small wavelength modes is suppressed compared to amplitude for large wavelength modes.
For wavelengths smaller than the 
co-moving free streaming scale for the quintessence 
fluid, $L_{fs}$, the fluctuations free-stream from over-dense to under-dense regions.
Thus, modes smaller than $L_{fs}$ experience oscillations and the damping of 
the power law growth of the inhomogeneous solution 
due to the competing effects of gravitational amplification and
pressure support from free streaming.  

In this section we obtained approximate analytic solutions to the 
fluctuation equation at long wavelengths. We used these solutions to explain 
the evolution of quintessence fluctuations for different equations of state. We found
that for $w$ closer to -1, the homogeneous solutions dominate the inhomogeneous
ones until later in the evolution of the universe.
In the next section we study the sensitivity of the CMB anisotropy to initial
conditions. We show that this longer lasting 
domination at values of the equation of state closer to -1
 determines the extent to which the initial 
conditions must be changed from the case of perfectly smooth initial conditions 
to affect the CMB anisotropy.

\section{Sensitivity to Initial Conditions}
\label{secinits}

\mynote{SENSITIVITY DEPENDS ON HOMOGENEOUS SOLNS}
We have seen in the last section 
that initial conditions affect only the 
homogeneous solutions of the fluctuation equation.
For models with $w$ closer to -1 such as $w=-2/3$ and $w=-0.9$, 
the homogeneous solution is larger and
dominates the inhomogeneous solution longer. In particular, the homogeneous solution
dominates at last scattering, and a change in initial conditions can propagate
forward in time to a change in the total fluctuation energy density, and consequently,
to a change in the temperature anisotropy.
Since the power law decline of the homogeneous solution is independent of wavelength, 
and the amplitude of the inhomogeneous solution is suppressed at smaller wavelengths,
any conclusions on sensitivity to initial conditions 
drawn at larger wavelengths will continue to hold at smaller ones.

\mynote{WHY RATIO OF ORDER UNITY}
At long wavelengths, an expression for the effect of the fluctuations in $\rho_m$ and 
$\rho_Q$ on the metric 
perturbation can be obtained in a very simple form from the Einstein 
equations \cite{MaandB}:
\be
\label{eqn.ree}
- {1\over 2}{\pri{a}\over a} \pri{h} \sim 4\pi G a^2 \delta\rho_m ( 2 + {\delta\rho_Q \over \delta\rho_m}).
\ee
The fluctuations in quintessence produce an effect which depends upon the ratio
$\delta \rho_Q / \delta \rho_{m}$. This ratio must become comparable to
unity at last scattering for there to 
be any distinguishable effect on the CMB anisotropy.
As can be seen in Figure \ref{k4comp}, in the case of adiabatic initial conditions, 
quintessence fluctuations are sub-dominant to matter fluctuations by many orders 
of magnitude for all $w$.
While the ratio $\delta \rho_Q / \delta \rho_{m}$ will increase if one
amplifies the initial conditions,
we will see that it is still too small at last scattering for most $w$ 
in order to have a distinguishable effect on the metric perturbation and 
consequently the CMB anisotropy.

\mynote{SMOOTH+ADIABATIC=INHOM SOLN+HOM SOLN}
To study the effect of changing initial conditions on the
CMB anisotropy, we start with the simplest possible initial conditions, smooth initial 
conditions, where the values of the fluctuation amplitudes 
$\delta\psi$ and $\delta\psi'$ 
are set to zero on the initial hyper-surface. Smooth initial conditions have the 
unique property that the quintessence fluctuation evolution
is determined solely by the inhomogeneous solution of the fluctuation equation. 
To test sensitivity, we compare to the case of adiabatic initial fluctuations in $Q$.
This corresponds to mixing the homogeneous solution into the inhomogeneous one.

\begin{figure}[t]
{\centering \begin{tabular}{lc}
\resizebox*{0.47\textwidth}{0.30\textheight}{\includegraphics{new_n9comp.eps}} &
\resizebox*{0.47\textwidth}{0.30\textheight}{\includegraphics{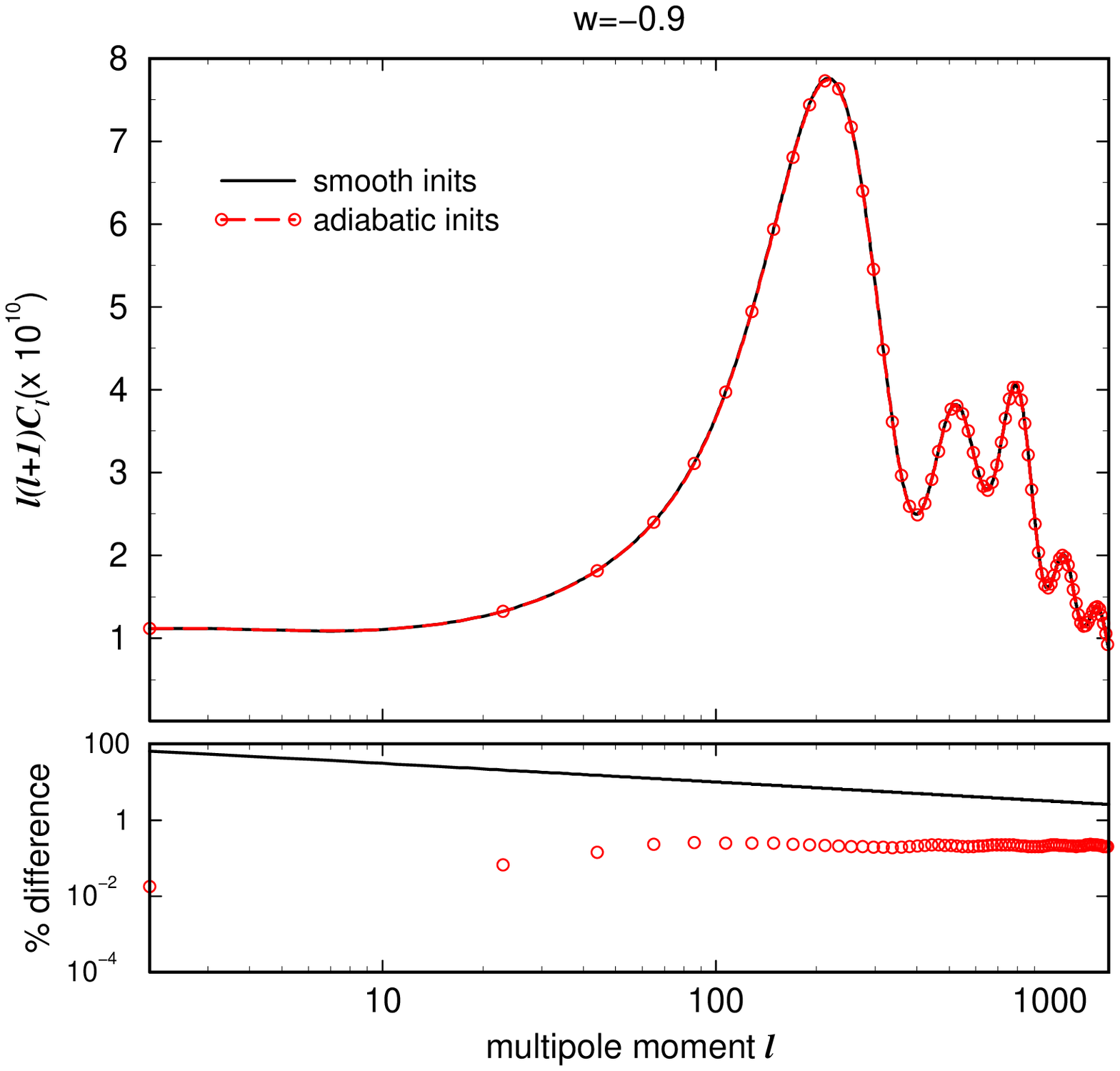}} \\
\end{tabular}}
\caption{
The figure in the left panel shows the evolution of the ratio of energy in $Q$
fluctuations to that in matter fluctuations
 $(\frac{\delta \rho_Q}{\delta \rho_M})^2$ at $k=10^{-4} Mpc^{-1}$ for
both smooth initial conditions (inhomogeneous 
solution) and adiabatic initial conditions 
at $w=-0.9$. The figure in the right panel 
shows the corresponding CMB power spectra as a function of multipole moment.
Plotted below the power spectrum 
is the percentage residual of the power spectrum for adiabatic 
initial conditions from smooth ones, compared to the fractional
cosmic variance ($100 \times {\Delta C_{\ell} \over C_{\ell}}$, plotted as a black line). 
The anisotropy change in going from smooth to adiabatic initial 
conditions is well below the variance
}
\label{new933comp}
\end{figure}

\mynote{SMOOTH TO ADIABATIC CHANGE MAKES NO DIFFERENCE TO CMB}
In the left panel of Figure \ref{new933comp}, we compare
the evolution of the ratio $\delta \rho_Q / \delta \rho_{m}$
at large wavelength ($k=10^{-4} Mpc^{-1}$) for smooth and adiabatic initial conditions, 
at $w=-0.9$. 
We see that the evolution of the ratio for the smooth case tracks 
the power law rise of the inhomogeneous solution to its present-day value.
The magnitude of the ratio at last scattering ($a \sim 10^{-3}$) 
is much larger in the adiabatic case than in the smooth case, 
corresponding to the dominance of the homogeneous solution over the inhomogeneous one.
Still, $\delta \rho_Q / \delta \rho_{m}$ 
remains far below unity in all epochs for the adiabatic case.

We plot the effect on the CMB anisotropy at $w=-0.9$ 
in the right panel of the figure, for 
both the smooth and adiabatic initial conditions. Below the
spectra is plotted the absolute value of 
the residual, or the percentage difference of the power spectrum in
the model with adiabatic initial conditions 
compared to the model with smooth initial conditions. We also plot
the fractional cosmic variance.
Residuals smaller than the variance cannot be observationally measured.
As can be seen from the figure, the anisotropy for the adiabatic case at $w=-0.9$ is 
not observationally distinguishable from the smooth case within cosmic variance.
While the addition of the homogeneous
solution to the inhomogeneous one does increase the ratio
$\delta \rho_Q / \delta \rho_{m}$ by many orders of magnitude, the 
increase is not enough, even at $w=-0.9$, to alter the CMB power spectrum.

\begin{figure}[t]
\centerline{\epsfxsize=3.3 in \epsfbox{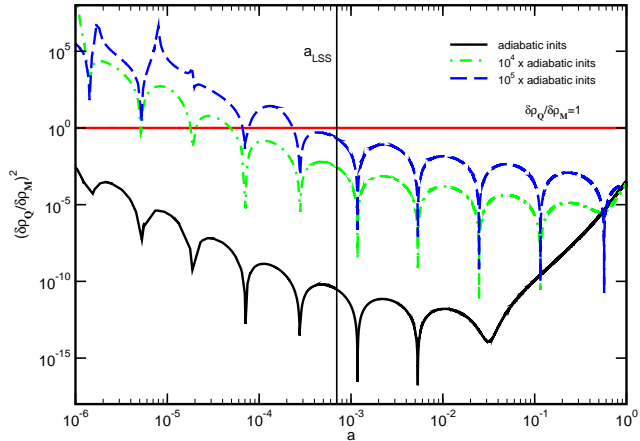}}
\caption{
The figure compares the evolution of the ratio 
 $(\frac{\delta \rho_Q}{\delta \rho_M})^2$ for
both adiabatic ($\eff = 1$) and artificially amplified ($\eff = 10^4$ and $\eff = 10^5$)
initial conditions for 
the $w=-0.9$ model from Figure \ref{k4comp}. The ratio is plotted for 
wave number $k=10^{-4} Mpc^{-1}$.  We also plot a solid horizontal line to indicate 
a ratio of magnitude unity. Notice that the amplification
prolongs the domination of the homogeneous solution, and the resultant 
closeness of the energy density ratio to unity. 
}
\label{rat4ws}
\end{figure}

\mynote{AMPLIFICATION BY $\eff$, EFFECT ON FLUCTS}
If there is to be a distinguishable imprint of a change in initial 
conditions on the CMB anisotropy, the energy density in quintessence fluctuations 
must increase drastically so that $\delta \rho_Q / \delta \rho_{m}$ 
is of order unity at last scattering. Let us estimate how large the
initial amplitude of $\delta \rho_Q/\rho_Q$ must be in order
to have a distinguishable effect by artificially multiplying
adiabatic initial conditions by a factor $\eff$, and then comparing
the result to the results for smooth initial conditions.  We want to
show that $\eff$ must be quite large in order to have any
detectable effect on the CMB anisotropy.
In Figure \ref{rat4ws} we display the evolution of 
$\delta \rho_Q / \delta \rho_{m}$ in the $w=-0.9$ model, for
initial conditions that are adiabatic ($\eff =1$), and for 
initial conditions $\eff=10^4$ and $\eff=10^5$ times the adiabatic initial conditions. 
The ratio is plotted at a wavelength longer than the horizon 
today, $k=10^{-4} Mpc^{-1}$.
In the case of amplified initial conditions, the homogeneous solution is larger,
and thus it takes until very recent epochs for the inhomogeneous solution 
to become comparable to the homogeneous solution. The 
amplification of the initial conditions by $\eff=10^4$ makes 
$\delta \rho_Q / \delta \rho_{m}$ larger than
unity initially, and the dominance of the homogeneous solution keeps it 
close to, but smaller than unity at last scattering.
An amplification by $\eff=10^5$ makes the ratio to be of order
unity at last scattering, which will leave an imprint on the CMB anisotropy. 

To understand these effects from the perspective of the value
of $\qfrac$, we need to obtain the value of the matter fluctuation 
at the initial hyper-surface ($a \sim 10^{-8}$). 
Since COBE normalization sets the amplitude of the matter fluctuation on re-entry to be 
$\mfrac \sim 10^{-5}$, we find from Eqn. \ref{eqn.thedbee} that 
$\Big(\mfrac\Big)_{init} \sim 10^{-16}$ in the
synchronous gauge. The ratio of $\Big(\qfrac\Big)_{init}$ to $\Big(\mfrac\Big)_{init}$ 
is set either by imposing smooth or adiabatic initial conditions. In the former case we have
$\delta \rho_Q/\rho_Q  = 0$ and so the ratio is zero.
In the latter case, the ratio can be obtained by 
combining Eqns. \ref{ic1} and \ref{ic2} with the 
scale-factor dependence of $\rho_m$ and $\rho_Q$.
The absolute value of this ratio ranges from $\sim 10^{-1}$ at $w=0$ to $10^{22}$ at $w=-0.9$. 
In other words, for $w$ closer to -1, $\qfrac$ is initially quite large.
Yet there is no observable difference in anisotropy, as can be seen from Figure. \ref{new933comp},
between the cases of smooth
and adiabatic initial conditions. The amplitude of
the homogeneous fluctuations swiftly declines and both the above ratio, and consequently
$\delta \rho_Q / \delta \rho_{m}$ are much smaller than one by last
scattering. Thus there is no observable change in the CMB anisotropy. 
It is only when the initial conditions are amplified 
by $F=10^5$ (so that $\Big(\qfrac\Big) \sim 10^{11}$ at the initial hyper-surface) 
that the steep decline cannot
offset the initially large value by the epoch of last scattering, and
there is any observable effect on the CMB.  Of course, this
large value of $\eff$ is physically unrealistic, many orders of magnitude
greater than what is expected from inflation, for example.  Also, for such
extreme values of $\eff$, the linear approximation used in CMB analysis is 
invalid.  This exercise shows clearly that we can ignore the initial
conditions on the quintessence fluctuations for all reasonable models.


\begin{figure}[t]
\centerline{\epsfxsize=3.3 in \epsfbox{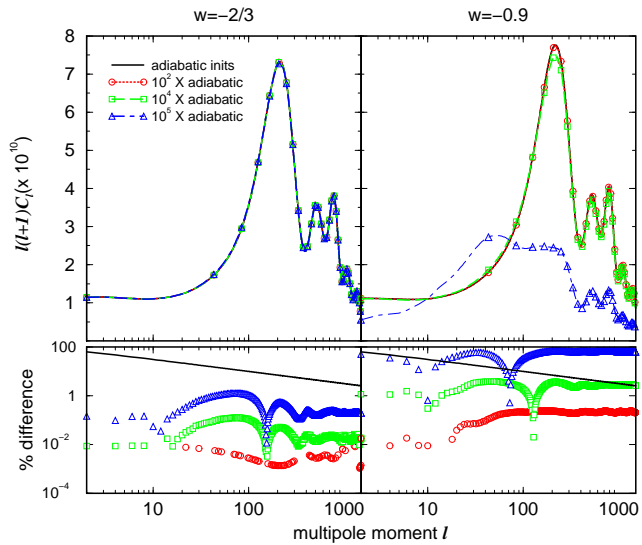}}
\caption{
The figure depicts the CMB power spectrum as a function of multipole moment
 for two of the models of  Figure \ref{k4comp} 
with $w=-2/3$ and $w=-0.9$. The power spectra are plotted
 for a series of cases with artificially amplified initial conditions, and for the
corresponding model with adiabatic initial conditions. Also shown in the lower panel
is the absolute value of the percentage residual of the amplified cases from the adiabatic case, as
compared to the fractional cosmic variance (black line).
At $w=-2/3$, an amplification of the adiabatic initial conditions even by
$\eff = 10^5$, is not enough to make an observable change in the CMB power spectrum.
On the other hand, in the $w=-0.9$ case, the power spectrum for the same value of $\eff$
is markedly different. 
}
\label{cmbgraph}
\end{figure}

\mynote{HOW MUCH AMPLIFICATION NEEDED TO AFFECT CMB}
In Figure \ref{cmbgraph}, we plot the numerically computed power spectra
at $w=-2/3$ and $w=-0.9$ for adiabatic and amplified
initial conditions. The amplification is by factors of $\eff = 10^2$,
 $10^4$ and $10^5$ times the adiabatic initial conditions.  Below the
spectra, we plot the residuals with respect to the adiabatic model.

We see that at $w=-2/3$, the power spectrum in the amplified
models is identical to that in the adiabatic model,
even for $\eff = 10^5$. For $w=-0.9$, 
an amplification by a factor of $10^2$ leads to no distinguishable 
changes in the anisotropy.
On the other hand, an amplification by $\eff=10^4$  weakly suppresses the Doppler
peak and creates changes in the anisotropy at some multipoles.  
The ratio of energy in quintessence fluctuations to energy 
in matter fluctuations for $\eff = 10^4$ is of the order $\sim 10^{-2}$-$10^{-1}$ at last 
scattering, as can be seen in Figure \ref{rat4ws}, and the residual
anisotropy is barely smaller than the cosmic variance.
Thus, the effects on the CMB are not large enough to be observationally distinguished.
Amplification of the initial conditions in the $w=-0.9$ case by a factor of $10^5$ 
raises the amplitude of the homogeneous solutions sufficiently
that ${\delta\rho_Q \over \delta\rho_m} \sim 1$ at last scattering. 
For equations of state even closer to -1, smaller amplification factors
are required to make a measurable difference in the CMB anisotropy. However, the 
amplification is still large and unphysical. For example, even at
$w=-0.999$, an amplification by $\eff = 10^3$ is required to create an observable
effect. 

\section{Conclusions}
\label{sec.concl}

We studied in this paper a large class of quintessence models with light fields and 
sound speed $c_s^2 \sim 1$ at small wavelengths, which have the property that they can be well 
approximated by constant equation of state, $w$. 
The evolution of the fluctuations in these models was
obtained by numerical integration and explained by approximate analytic solutions to the
fluctuation equation at large wavelengths.
Our central result is that the CMB anisotropy
in such models is insensitive to initial conditions on the quintessence fluctuations 
for smooth and adiabatic initial conditions . For $w=-0.9$, the CMB anisotropy 
is insensitive in the large range of initial conditions 
$\Big(\qfrac\Big)_{init} < 10^{11} \,(F=10^5)$ 
for $\Big(\mfrac\Big) \sim 10^{-5}$ at horizon re-entry.
Secondly, the sensitivity increases as $w$ approaches -1. At $w=-0.999$, the range reduces to
$\Big(\qfrac\Big)_{init} < 10^{9} \,(F=10^3)$. However, physically reasonable models 
such as those based on inflation and ekpyrosis do not produce such large values of $\qfrac$, 
and the ratio of energy in quintessence fluctuations 
to that in matter fluctuations is much smaller
than unity. Hence, we do not anticipate
that the CMB anisotropy will be sensitive to initial conditions in realistic cases. The same 
analytical arguments made in this paper carry over to the more general quintessence models in
which $w$ is more strongly time dependent or $c_s^2 \ne 1$ at small wavelengths. However, the
precise numerical lower bound on the initial conditions required to imprint
a distinguishable effect on the CMB anisotropy has to be worked out on a case by case basis.

\medskip
 
 This work was supported by
 DOE grant DE-FG02-95ER40893 (RD), NSF grant PHY-0099543 (RRC),
  and DOE grant DE-FG02-91ER40671 (PJS).

\noindent

\end{document}